\documentclass[preprint,12pt]{elsarticle}

\usepackage{amssymb}
\usepackage{amsmath}

\usepackage{tikz}
\usepackage[compat=1.1.0]{tikz-feynhand}
\usepackage{tikz-feynman}

\journal{Physics Letters A}

\begin{document}

\begin{frontmatter}



\title{Quantum second harmonic generation in terms of elementary processes}


\author[inst1]{Giovanni Chesi}

\affiliation[inst1]{organization={National Institute for Nuclear Physics, Sezione di Pavia},
            addressline={Via Agostino Bassi 6}, 
            city={Pavia},
            postcode={27100}, 
            state={Italy}}


\begin{abstract}
We address the quantum dynamics of second harmonic generation with a perturbative approach. By inspecting the Taylor expansion of the unitary evolution, we identify the subsequent application of annihilation and creation operators as elementary processes and find out how the expansion of the second-harmonic photon-number probability distribution can be expressed in terms of the interplay of these processes. We show that overlaps between the output states of different elementary processes contribute to the expansion of the probability distribution and provide a diagrammatic technique to analytically retrieve terms of the distribution expansion at any order. 
\end{abstract}




\end{frontmatter}


\section{Introduction}
\label{intro}
Second harmonic generation (SHG) was the first optical nonlinear phenomenon experimentally observed \cite{Franken1961} and one of the first for which a complete classical description was provided \cite{Armstrong1962}. Since then, much effort has been devoted to increase the SHG efficiency \cite{Zhu2004,Dimova2018,Frigerio2018,Franchi2020,Liu2021,Yuan2021,Frigerio2021,DeLeonardis2022,Chesi2023}. Interestingly, achieving giant nonlinear couplings may be promising to exploit SHG as a source of nonclassical light. Indeed, SHG can be used to generate squeezed states, both on the pump and on the second-harmonic (SH) mode \cite{Mandel1982,Kielich1987,Hillery1987,Zhan1991}. The nonclassicality of the output light in SHG has been experimentally assessed \cite{Huang1992}, remarkably in terms of sub-Poissonian statistics \cite{Youn1996}. However, an exact prediction of the quantum SHG dynamics cannot be provided since the pertaining evolution operator cannot be disentangled in a closed form. Indeed, the interaction Hamiltonian describing SHG is not linear nor bilinear in the field modes. It reads
\begin{equation} \label{ham}
    H = \gamma [a_1^{2}a_2^{\dagger} + (a_1^{\dagger})^2a_2]
\end{equation}
where $\gamma$ is the coupling with the nonlinear material, $a_X$ and $a_X^{\dagger}$ are the annihilation and creation operators, respectively, for the pump mode ($X=1$) and for the SH mode ($X=2$), with the usual commutation relations $[a_X,a^{\dagger}_Y] = \delta_{X,Y}$. Note that here we included the Planck constant in the coupling $\gamma$. The Hamiltonian of the free fields, given by

\begin{equation}
    H_0 = \omega a_1^{\dagger}a_1 + 2\omega a_2^{\dagger}a_2,
\end{equation}
is a constant of motion. Therefore, the interaction Hamiltonian $H$ in Eq.~(\ref{ham}) is a constant of motion too, as pointed out in Ref.~\cite{Mostowski1978}. Moreover, we remark that $H_0$ and $H$ commute, namely $[H_0,H]=0$, which implies that $H$ does not depend on time also in the interaction picture. Then, the evolution of our SHG system is simply described by the unitary $U$ generated by the interaction Hamiltonian $H$, namely $U = e^{iHt}$. Here we will include also the time dependence in the coupling $\gamma$, i.e. we will replace $\gamma t$ with $\gamma(t)\equiv \gamma t$.

Many strategies have been tried to retrieve the dynamics of SHG and many results obtained. In 1967, the density matrix of the up-converted field was derived in Ref.~\cite{Shen1967} by exploiting the parametric approximation, i.e. by assuming that the pump field operators are $c$-numbers, due to the intensity of the field. In Ref.~\cite{Bonifacio1969}, it was proved the equivalence between a system of $N$ two-level atoms interacting with an electromagnetic field, described by the Dicke model, and a generic trilinear-boson-scattering process; hence, in Ref.~\cite{Orszag1983}, the photon-number generation probability and the efficiency of the process were retrieved. The equivalence is essentially based on the rotating-wave approximation, but, again, the analytic results rely on a complete factorization of the bosonic operators. 
\\
In Refs.~\cite{Crosignani1972,Chmela1973}, the equation of motion for the expectation value of the pump photon-number operator was retrieved and solved for coherent and chaotic input states.
\\
The linearization method, developed in Refs.~\cite{Li1993,Li1994}, was used to investigate the evolution of the quantum noise in traveling-wave SHG under the assumption that the quantum effects are small perturbations with respect to the mean values of the fields. 
\\
For long interaction times, numerical methods were exploited in Ref.~\cite{Bajer1999} to retrieve the behavior of the output light statistics in the case of degenerate three-wave mixing.
\\
Much attention has been devoted to perturbation methods. In Refs.~\cite{Tanas1973, Kozierowski1977,Ekert1988}, the first- and second-order Glauber's correlation functions were expanded up to the fourth order through the derivatives of the annihilation and creation operators of the pump and SH field. These results eventually led to find in Refs.~\cite{Mandel1982,Kielich1987,Hillery1987,Zhan1991} the squeezing of the fundamental and SH field quadratures. In Ref.~\cite{Bajer1992}, the perturbation approach was conceived as a Taylor expansion of the annihilation and creation operators evolved by the unitary $U$ and the orders of the series were numerically evaluated in the case of a coherent input state; the results assessed the sub-Poissonian photon-number statistics of the output SH field. Recently, we exploited a similar method to determine the SH photon-number statistics for different input pump states, namely Fock, coherent, chaotic and squeezed states \cite{Chesi2019}. In particular, we expanded both the output SH state and the evolved SH number operator analytically up to the sixth order. The otuput state determined through the Taylor expansion of the unitary evolution $U$ results from the interplay between sequential applications of creation and annihilation operators. Interstingly, we found that different sequences of $a_X$ and $a^{\dagger}_X$ operators overlap in the perturbative expansion. The structure of these contributions determines the quantum effects in the output SH photon-number statistics, such as the sub-Poissonian behavior.

Here we identify these sequences of creation and annihilation operators as elementary processes, clarify their structure and role in the perturbative approach and provide a consistent description of their superpositions in the case of SHG. We show that the expansion of the SH photon-number distribution can be expressed entirely in terms of superposition of elementary processes. This analysis allows us to investigate the physical meaning of the terms in the perturbative expansion, but also yields a technique to visualize and compute them at any order. In this regard, we introduce simplified double-sided Feynman diagrams that uniquely identify superpositions of processes and show how one can compute the pertaining contribution to the expansion of the SH photon-number distribution at any order.  

The paper is structured as follows. In Sect.~\ref{2} we define the elementary processes and the process superposition in the case of spontaneous SHG. We show how the perturbative expansion of the SH photon-number distribution can be expressed in terms of superposition of elementary processes.
In Sect.~\ref{3} we develop a diagrammatic approach that allows one to better visualize the contribution of the elementary processes to the SH photon-number distribution. In particular, in Sect.~\ref{31} we simplify the double-sided Feynman diagrams for spontaneous SHG. In Sect.~\ref{32} we show how these diagrams can be used to retrieve perturbative terms of the SH photon-number distribution at any order.
Finally, in Sect.~\ref{4}, we draw our conclusions.


\section{Elementary processes and process superposition}
\label{2}
In this paper we will focus on spontaneous SHG \cite{Bajer1999}, which is also usually known as just SHG \cite{Bajer1992}: at the entrance of the nonlinear medium we have a fundamental-frequency beam with non-zero mean intensity and the SH beam in the vacuum state. Namely, the input state reads
\begin{equation} \label{input}
    \rho_0 = \sum_{n,m}c_{n,m}|n,0\rangle\langle m,0|
\end{equation}
where $|n\rangle$ and $|m\rangle$ are eigenstates of the number operator $n \equiv a^{\dagger}a$ and $c_{n,m}$ the coefficients of the fundamental-mode input state. The propagation inside the nonlinear medium is described by the unitary evolution $U$ with the Hamiltonian in Eq.~(\ref{ham}) and eventually generates the output state
\begin{equation} \label{output}
    \rho (\gamma) = U\rho_0 U^{\dagger} = \sum_{n,m}c_{n,m}U|n,0\rangle\langle m,0|U^{\dagger}.
\end{equation}
Note that we just need to evolve the Fock states $|n,0\rangle\langle m,0|$ and then the general case of a generic pump input state is recovered by summing over $n$ and $m$ with the weights $c_{n,m}$. To this aim, we expand the unitary evolution as follows
\begin{equation} \label{uex}
    U = \exp[i\gamma(a_1^2a_2^{\dagger}+(a_1^{\dagger})^2a_2)] \sim I + \sum_r \frac{(i\gamma)^r}{r!}[a_1^2a_2^{\dagger}+(a_1^{\dagger})^2a_2]^r
\end{equation}
where $I$ is the identity and, as mentioned in the Introduction, we included the time dependence in the coupling, namely $\gamma = \gamma(t) \equiv \gamma t$. The expansion in Eq.~(\ref{uex}) applied to the input state as in Eq.~(\ref{output}) allows us to clarify the effect of the unitary evolution by inspecting the physical meaning of the terms in the expansion. We will use our analysis to retrieve and inspect the related Taylor series for the SH photon-number probability distribution, namely
\begin{equation}
    \text{Pr}(n',v';\gamma) = \langle n',v'|\rho(\gamma)|n',v'\rangle.
\end{equation}

We define an \textit{elementary process} as any subsequent application of creation and annihilation operators. The sequence of operators does not need to be normally ordered. Note from the structure of the Hamiltonian that the sequence of operators $a_1$ and $a_1^{\dagger}$ is strictly related to the sequence of $a_2^{\dagger}$ and $a_2$, respectively, since for each photon creation on the SH mode we have the annihilation of two photons on the fundamental mode and vice versa. An elementary process can be formalized through an operator on the space spanned by the Fock states $|n,m\rangle$ as follows
\begin{equation}
    A_{\mathbf{k}} = (a_1^{\dagger})^{2k_l}...a_1^{2k_3}(a_1^{\dagger})^{2k_2}a_1^{2k_1}a_2^{k_l}...(a_2^{\dagger})^{k_3}a_2^{k_2}(a_2^{\dagger})^{k_1}
\end{equation}
where $\mathbf{k}\equiv \{k_j\}_{j=1}^l$ is the vector of natural numbers $k_j$. Note that at the order $r$ in the expansion of Eq.~(\ref{uex}) all the terms can be expressed in this form, with $\sum_{j=1}^lk_j=r$. Importantly, in the expansion of $U$ we find different elementary processes that map same input states into same output states, i.e., for $\mathbf{k} \neq \mathbf{k'}$,
\begin{equation} \label{prop}
    \begin{aligned}
        &A_{\mathbf{k}}|n,0\rangle = f_{\mathbf{k}}(n,v)|n-2v, v\rangle \\
        &A_{\mathbf{k'}}|n,0\rangle = f_{\mathbf{k'}}(n,v)|n-2v, v\rangle
    \end{aligned}
\end{equation}
with $f_{\mathbf{k}}$ and $f_{\mathbf{k'}}$ real functions of $n$ and $v$. Hence, when the expansion of $U$ is applied to the input state as in Eq.~(\ref{output}), at a generic order $r$ we find terms given by
\begin{equation}
    \frac{(i\gamma)^r}{r!}A_{\mathbf{k}}|n,0\rangle\langle m,0|A^{\dagger}_{\mathbf{k'}} = \frac{(i\gamma)^r}{r!}f_{\mathbf{k}}(n,v)f_{\mathbf{k'}}(m,v)|n-2v, v\rangle\langle m-2v,v|.
\end{equation}
Then, these terms identified by different elementary processes give to the SH photon-number distribution $\text{Pr}(n',v';\gamma)$ a contribution proportional to
\begin{equation} \label{supterms}
    \langle n',v'| A_{\mathbf{k}}|n,0\rangle\langle m,0|A^{\dagger}_{\mathbf{k'}}|n',v'\rangle = f_{\mathbf{k}}(n,v)f_{\mathbf{k'}}(n,v)\delta_{n',n-2v}\delta_{v',v}
\end{equation}
which in general is not null. In order to better clarify these contributions, let us define a suitable map that exploits the property outlined in Eqs.~(\ref{prop}), i.e. given a same input Fock state for two different elementary processes, the output states have non-zero overlap. Consider the product states of a Fock space $\sigma\equiv|n, v\rangle\langle m,w|$, $\tilde{\sigma}\equiv|\tilde{n}, \tilde{v}\rangle\langle \tilde{m},\tilde{w}|$ and two elementary processes $A_{\mathbf{k}}$, $A_{\mathbf{k'}}$. We define \textit{process superposition} the automorphism $\alpha_{\mathbf{k},\mathbf{k'}}$ such that
\begin{equation}
  \alpha_{\mathbf{k},\mathbf{k'}}(\sigma)  \equiv A_{\mathbf{k}}\sigma A_{\mathbf{k'}} = a\,\tilde{\sigma}
\end{equation}
where $a$ is a real number. The expansion of the SH photon-number probability distribution is a sum over $\mathbf{k}$ and $\mathbf{k'}$ (i.e. over all the elementary processes) of the non-null diagonal element $s_{\mathbf{k},\mathbf{k'}}$ of the process-superposition maps $\alpha_{\mathbf{k},\mathbf{k'}}$, namely
\begin{equation}
\begin{aligned}
    s_{\mathbf{k},\mathbf{k'}}(n',v'|n,0) &\equiv \langle n',v'| \alpha_{\mathbf{k},\mathbf{k'}}(|n,0\rangle\langle m,0|)|n',v'\rangle \\
    &= \langle n',v'| \alpha_{\mathbf{k},\mathbf{k'}}(|n,0\rangle\langle n,0|)|n',v'\rangle.
\end{aligned}
\end{equation}

Note that, if $\mathbf{k} = \mathbf{k'}$, the term in Eq.~(\ref{supterms}) reduces to the photon-number probability for a single elementary process, i.e.
\begin{equation}
\begin{aligned}
    s_{\mathbf{k},\mathbf{k}}(n',v'|n,0) &= f(n,v)f(m,v)\langle n',v'| n-2v,v\rangle\langle m-2v,v|n',v'\rangle \\
    &= f(n,v)f(m,v) \delta_{n,m}\delta_{n',n-2v}\delta_{v',v} \\ &= f^2(n,v)\delta_{n',n-2v}\delta_{v',v} \\
    &= |\langle n',v'| A_{\mathbf{k}}|n,0\rangle|^2 \equiv p_{\mathbf{k}}(n',v'|n,0).
\end{aligned}
\end{equation}
Ultimately, the SH photon-number distribution can be expressed as the following expansion
\begin{equation} \label{probexp}
    \begin{aligned}
        \text{Pr}(n',v';\gamma) &\sim \sum_n c_{n,n} \sum_{r,r'}\frac{(i\gamma)^r}{r!}\frac{(-i\gamma)^{r'}}{r'!}\sum_{\mathbf{k},\mathbf{k'}}s^{(r,r')}_{\mathbf{k},\mathbf{k'}}(n',v'|n,0) \\
        &= \sum_n c_{n,n} \sum_{r,r'}\frac{(i\gamma)^r}{r!}\frac{(-i\gamma)^{r'}}{r'!}\sum_{\mathbf{k},\mathbf{k'}}f_{\mathbf{k}}^{(r)}(n,v)f_{\mathbf{k'}}^{(r')}(n,v) \delta_{n',n-2v}\delta_{v',v}
    \end{aligned}
\end{equation}
where the matrix elements $s^{(r,r')}_{\mathbf{k},\mathbf{k'}}(n',v'|n,0)$ are specified for the elementary processes $A_{\mathbf{k}}^{(r)}$ and $A_{\mathbf{k'}}^{(r')}$ at the orders $r$ and $r'$, i.e. with $\sum_{j=1}^lk_j = r$ and $\sum_{j=1}^{l'}k'_j = r'$, respectively. Namely,
\begin{equation}
    s^{(r,r')}_{\mathbf{k},\mathbf{k'}}(n',v'|n,0) = \langle n',v'| A_{\mathbf{k}}^{(r)}|n,0\rangle\langle n,0|(A_{\mathbf{k'}}^{\dagger})^{(r')}|n',v'\rangle.
\end{equation}
Therefore, one can approximate the SH photon-number distribution (and, hence, its pertaining moments) up to a given order $R = r+r'$ by evaluating the non-null diagonal elements of the matrix $\sum_{\mathbf{k},\mathbf{k'}}\alpha_{\mathbf{k},\mathbf{k'}}(|n,0\rangle\langle n,0|)$ up to $R$. In the next Section, we will show a diagrammatic approach to visualize any process superposition and compute the corresponding contribution for any order of the expansion.


\section{Diagrammatic approach}
\label{3}
\subsection{Double-sided Feynman diagrams re-visited for spontaneous SHG}
\label{31}
In order to evaluate the contributions of the process superpositions to the SH photon-number distribution for a given order of its expansion in Eq.~(\ref{probexp}), we exploit a re-visited version of the double-sided Feynman diagrams \cite{Yee1978,Prior1984,Kumar1992,Boyd2008}. 

Each of our diagrams describes a process superposition, namely
\begin{equation}
\alpha^{(r,r')}_{\mathbf{k},\mathbf{k'}}(|n,v\rangle\langle m,w|) \equiv A_{\mathbf{k}}^{(r)}|n,v\rangle\langle m,w|(A_{\mathbf{k'}}^{\dagger})^{(r')}.
\end{equation}
Since here we focus on the contributions to the SH photon-number distribution for spontaneous SHG, we will just consider the diagrams related to process superpositions given by $\alpha^{(r,r')}_{\mathbf{k},\mathbf{k'}}(|n,0\rangle\langle n,0|)$, at the order $R=r+r'$ of the distribution expansion.

The diagrams are double-sided because they display two sides, one on the left and one on the right. The left side represents the elementary process $A_{\mathbf{k}}^{(r)}$ acting on the ket state $|n,0\rangle$, the right side the elementary process $(A_{\mathbf{k'}}^{\dagger})^{(r')}$ acting on the bra state $\langle n,0|$.

The electromagnetic-field state $|n-2v,v\rangle$, with $n$ number of input pump photons and $v$ number of SH photons created, namely the tensor product between the Fock state of the fundamental mode $|n-2v\rangle$ and the Fock state of the SH mode $|v\rangle$, is represented as a wiggly line with an arrow:
\begin{tikzpicture}
\begin{feynhand}
\vertex (a) at (0,0);      
\vertex (b) at (1,0) {};
\propag [chabos] (a) to (b);
\end{feynhand}
\end{tikzpicture}.

The atomic field in the ground state $|g\rangle$ is represented by a plain line with an arrow: \begin{tikzpicture}
\begin{feynhand}
\vertex (a) at (0,0);      
\vertex (b) at (1,0) {};
\propag [fer] (a) to (b);
\end{feynhand}
\end{tikzpicture}. Note that the unitary evolution $U$ identified by the Hamiltonian in Eq.~(\ref{ham}) leaves unchanged the atomic state of the nonlinear medium, whose role here is just to couple the electromagnetic fields of the pump and SH mode. Therefore, in our diagrams the whole evolution of the atomic state will be implied in the interactions, which will always take the ground state into the ground state.

The interaction between the electromagnetic modes and the atomic field is represented by a vertex: 
\begin{tikzpicture}
\begin{feynhand}
\vertex [dot] (a) at (0,0) {};      
\end{feynhand}
\end{tikzpicture}. Each interaction is one of the creation or annihilation operators of the elementary process described by one side of the diagram. The sequence of the operators is oriented from the bottom to the top. If in the sequence we have two or more consecutive creation or annihilation operators on the same mode, i.e. $(a_2^{\dagger})^{k_j}$ or $a_2^{k_j}$, respectively, then we reduce the sequence of identical operators to just one vertex with the number of its repetitions reported aside, namely: 
\begin{tikzpicture}
\begin{feynhand}    
\vertex [dot] (a) {};
\end{feynhand}
\end{tikzpicture} $k_j$. In the case of input and output fields, for the ket states (left side) the arrow points towards the vertex, while for bra states (right side) it points away from the vertex. If a field connects two vertices or if there are no vertices, the direction of the arrow follows the temporal order (from the bottom to the top) for ket states and the other way for the bra states.

As an example, we show here the diagrams of the process superpositions that give a non-null contribution at the second order of the distribution expansion in Eq.~(\ref{probexp}).
\\
\\
\\
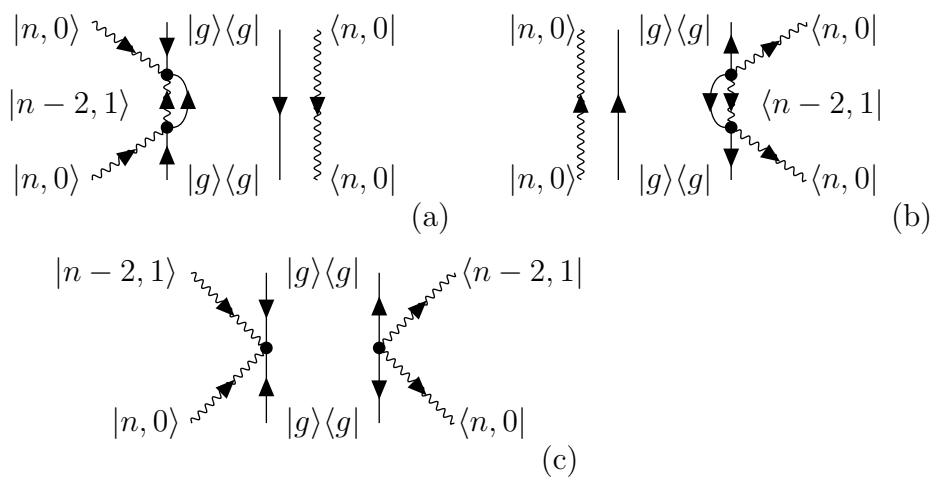
\begin{tikzpicture}
\begin{feynhand}    
\vertex (a) at (0,0);      
\vertex (b) at (1,0);
\vertex [dot] (c) at (1,0.7) {};   
\vertex [dot] (d) at (1,1.4) {};
\vertex (e) at (1,2.1);
\vertex (f) at (0,2.1);
\propag [chabos] (a) to (c);
\propag [fer] (b) to (c);
\propag [chabos] (c) to (d);
\propag [fer] (c) to [out=0, in=0] (d);
\propag [fer] (e) to (d);
\propag [chabos] (f) to (d);
\node at (-0.6,2) {$|n,0\rangle$}; 
\node at (-0.3,1) {$|n-2,1\rangle$};
\node at (-0.6,0) {$|n,0\rangle$};
\node at (1.5,0) {$|g\rangle$}; 
\node at (1.5,2) {$|g\rangle$}; 
\vertex (a') at (2.5,0);      
\vertex (b') at (3,0);
\vertex (c') at (2.5,2);   
\vertex (d') at (3,2);
\propag [fer] (c') to (a');
\propag [chabos] (d') to (b');
\node at (2,0) {$\langle g|$};
\node at (2,2) {$\langle g|$};
\node at (3.6,0) {$\langle n,0|$};
\node at (3.6,2) {$\langle n,0|$};
\node at (4.5,-0.5) {(a)};

\vertex (a') at (6.5,0);      
\vertex (b') at (7,0);
\vertex (c') at (6.5,2);   
\vertex (d') at (7,2);
\propag [chabos] (a') to (c');
\propag [fer] (b') to (d');
\node at (7.5,0) {$|g\rangle$};
\node at (7.5,2) {$|g\rangle$};
\node at (6,0) {$|n,0\rangle$};
\node at (6,2) {$|n,0\rangle$};
\vertex (a) at (8.5,0);      
\vertex (b) at (9.5,0);
\vertex [dot] (c) at (8.5,0.7) {};   
\vertex [dot] (d) at (8.5,1.4) {};
\vertex (e) at (8.5,2.1);
\vertex (f) at (9.5,2.1);
\propag [fer] (c) to (a);
\propag [chabos] (c) to (b);
\propag [chabos] (d) to (c);
\propag [fer] (d) to [out=180, in=180] (c);
\propag [chabos] (d) to (f);
\propag [fer] (d) to (e);
\node at (8,2) {$\langle g|$}; 
\node at (8,0) {$\langle g|$};
\node at (10,2) {$\langle n,0|$}; 
\node at (9.7,1) {$\langle n-2,1|$}; 
\node at (10,0) {$\langle n,0|$}; 
\node at (10.9,-0.5) {(b)};
\end{feynhand}
\end{tikzpicture}

\begin{tikzpicture}
\begin{feynhand}    
\vertex (a) at (0,0);      
\vertex (b) at (1,0);
\vertex [dot] (c) at (1,1) {};   
\vertex (d) at (0,2);
\vertex (e) at (1,2);
\propag [chabos] (a) to (c);
\propag [fer] (b) to (c);
\propag [fer] (e) to (c);
\propag [chabos] (d) to (c);
\node at (-1,2) {$|n-2,1\rangle$}; 
\node at (-0.6,0) {$|n,0\rangle$};
\node at (1.5,0) {$|g\rangle$}; 
\node at (1.5,2) {$|g\rangle$}; 
\vertex (a') at (2.5,0);      
\vertex (b') at (3.5,0);
\vertex [dot] (c') at (2.5,1) {};   
\vertex (d') at (3.5,2);
\vertex (e') at (2.5,2);
\propag [fer] (c') to (a');
\propag [chabos] (c') to (b');
\propag [fer] (c') to (e');
\propag [chabos] (c') to (d');
\node at (2,2) {$\langle g|$}; 
\node at (2,0) {$\langle g|$};
\node at (4.4,2) {$\langle n-2,1|$}; 
\node at (4,0) {$\langle n,0|$}; 
\node at (4.9,-0.5) {(c)};
\end{feynhand}
\end{tikzpicture}

Diagrams (a) and (b) show the superposition $(a_1^{\dagger})^2a_1^2a_2a_2^{\dagger}|n,0\rangle\langle n,0|I$ and its complex conjugate, namely the superposition between the subsequent creation and annihilation of one SH photon with the no-interaction process. Diagram (c) shows the creation of one SH photon out of the elementary process $a_1^2a_2^{\dagger}$. This last diagram specifies in the distribution expansion of Eq.~(\ref{probexp}) the probability that one SH photon is created through the elementary process $a_1^2a_2^{\dagger}$. In particular, the second order $\text{Pr}^{(2)}$ of the SH photon-number distribution reads
\begin{equation}
    \begin{aligned}
        &\text{Pr}^{(2)}(n',v';\gamma) = \\
        =&  \sum_n c_{n,n}\langle n',v'|\frac{(i\gamma)^2}{2!}(a_1^{\dagger})^2a_1^2a_2a_2^{\dagger}|n,0\rangle\langle n,0|I + \frac{(-i\gamma)^2}{2!}I|n,0\rangle\langle n,0|a_2a_2^{\dagger}(a_1^{\dagger})^2a_1^2 \\
        &+ (i\gamma)(-i\gamma)a_1^2a_2^{\dagger}|n,0\rangle\langle n,0|a_2(a_1^{\dagger})^2 |n',v'\rangle  \\
        =& \sum_n c_{n,n} \gamma^2n(n-1)(\delta_{n',n}\delta_{v',0} + \delta_{n',n-2}\delta_{v',1}).
    \end{aligned}
\end{equation}
The first two terms of the second equality are given by the process superpositions represented by diagrams (a) and (b). Note that they provide the same contribution in terms of output state and probability amplitudes, so that they sum to each other. This is general: every process superposition with $\mathbf{k}\neq\mathbf{k'}$ provide the same contribution as its complex conjugate.

Two second-order process superpositions that do not contribute to the second-order distribution expansion are the following ones.
\\
\\
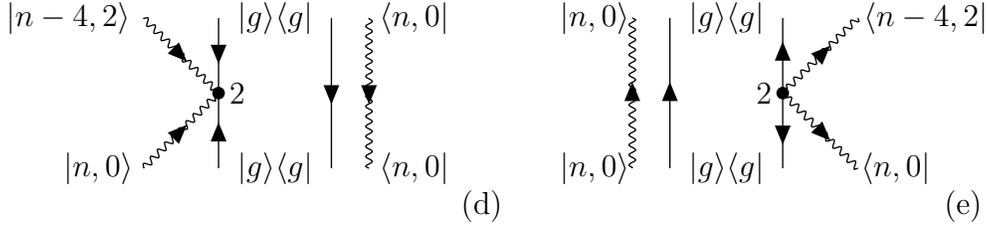
\begin{tikzpicture}
\begin{feynhand}    
\vertex (a) at (0,0);      
\vertex (b) at (1,0);
\vertex [dot] (c) at (1,1) {};   
\vertex (d) at (0,2);
\vertex (e) at (1,2);
\propag [chabos] (a) to (c);
\propag [fer] (b) to (c);
\propag [fer] (e) to (c);
\propag [chabos] (d) to (c);
\node at (-1,2) {$|n-4,2\rangle$}; 
\node at (-0.6,0) {$|n,0\rangle$};
\node at (1.5,0) {$|g\rangle$}; 
\node at (1.5,2) {$|g\rangle$}; 
\node at (1.25,1) {2}; 
\vertex (a') at (2.5,0);      
\vertex (b') at (3,0);
\vertex (c') at (2.5,2);   
\vertex (d') at (3,2);
\propag [fer] (c') to (a');
\propag [chabos] (d') to (b');
\node at (2,0) {$\langle g|$};
\node at (2,2) {$\langle g|$};
\node at (3.6,0) {$\langle n,0|$};
\node at (3.6,2) {$\langle n,0|$};
\node at (4.5,-0.5) {(d)};

\vertex (a') at (6.5,0);      
\vertex (b') at (7,0);
\vertex (c') at (6.5,2);   
\vertex (d') at (7,2);
\propag [chabos] (a') to (c');
\propag [fer] (b') to (d');
\node at (7.5,0) {$|g\rangle$};
\node at (7.5,2) {$|g\rangle$};
\node at (6,0) {$|n,0\rangle$};
\node at (6,2) {$|n,0\rangle$};
\vertex (a) at (8.5,0);      
\vertex (b) at (9.5,0);
\vertex [dot] (c) at (8.5,1) {};   
\vertex (d) at (9.5,2);
\vertex (e) at (8.5,2);
\propag [fer] (c) to (a);
\propag [chabos] (c) to (b);
\propag [fer] (c) to (e);
\propag [chabos] (c) to (d);
\node at (8,2) {$\langle g|$}; 
\node at (8,0) {$\langle g|$};
\node at (10.4,2) {$\langle n-4,2|$}; 
\node at (10,0) {$\langle n,0|$}; 
\node at (8.25,1) {2}; 
\node at (10.9,-0.5) {(e)};
\end{feynhand}
\end{tikzpicture}

Indeed, note that here the two overlapping processes do not output the same state.

\subsection{From diagrams to the SH photon-number distribution}
\label{32}
Now we show how to compute from a diagram the corresponding term of the SH photon-number probability expansion in Eq.~(\ref{probexp}). 
\\
A diagram contributes to the distribution expansion only if for both the sides
\begin{itemize}
    \item the first interaction is a creation operator,
    \item $\forall j$ the number of creation operators at vertex $j$ is equal or larger than the number of annihilation operators at vertex $j+1$,
    \item the output states are the same under complex conjugation.
\end{itemize}
If the two sides of the diagram display the same elementary process $A_{\mathbf{k}}$ (i.e. if the sequence of annihilation and creation operators is the same, $\mathbf{k} = \mathbf{k'}$), then the diagram represents the probability of generating photons through that elementary process. Then, one can perform the operations that follow just for one side of the diagram and then compute the square modulus. If $\mathbf{k} \neq \mathbf{k'}$, the diagram represents a process superposition, thus its complex conjugate provides an identical contribution to the distribution. Therefore, when writing the sum of all the diagrams for a given order of the distribution expansion, one can neglect the latter and multiply by two the result obtained from the former.
\\
The perturbative order for which a diagram gives a contribution can be found as follows. Each side provides the perturbative order of the corresponding unitary expansion, which reads $\sum_jk_j \equiv r$ for the elementary process $A_{\mathbf{k}}$ on the left side and $\sum_jk_j' \equiv r'$ for the elementary process $A_{\mathbf{k'}}$ on the right. Then, the order of the diagram is $R\equiv r+r'$. Note that $R$ can only be an even number, since $r$ and $r'$ must be both even or both odd for the two sides to generate the same number of SH photons. We associate to the diagram a factor 
\begin{equation}
    \frac{i^r(-i)^{r'}}{r!r'!}\gamma^R.
\end{equation}
\\
Now we focus on the vertices on one side of the diagram. As mentioned above, for the left (right) side, the $j$-th vertex from the bottom represents the application of the same operator, either $a_1^2a_2^{\dagger}$ or $(a_1^{\dagger})^2a_2$, for $k_j$ ($k_j'$) times, with $k_j \in [0,r]$ ($k_j' \in [0,r']$). Let us consider the left side of the diagram. If the $j$-th vertex of the diagram represents the creation of $k_j$ SH photons and the annihilation of $2k_j$ pump photons, namely the operator $a_1^{2k_j}(a_2^{\dagger})^{k_j}$, then $j$ is an odd number and $j\geq 1$. We associate to this vertex a second-harmonic-creation (shc) factor

\begin{equation} \label{fshc}
    f_{\text{shc}}(n,k_j) \equiv \sqrt{\frac{K_j!}{(K_j - k_j)!}\frac{[n-2(K_j - k_j)]!}{(n - 2K_j)!}}
\end{equation}
where
\begin{equation} \label{KJ}
    K_j \equiv \sum_{q=1}^j(-1)^{q-1}k_q.
\end{equation}
If, on the contrary, the $j$-th vertex of the diagram represents the annihilation of $k_j$ SH photons and the creation of $2k_j$ pump photons, namely the operator $(a_1^{\dagger})^{2k_j}a_2^{k_j}$, then $j$ is an even number and $j\geq 2$. We associate to this vertex a second-harmonic-annihilation (sha) factor

\begin{equation} \label{fsha}
    f_{\text{sha}}(n,k_j) \equiv \sqrt{\frac{(K_j+k_j)!}{K_j!}\frac{(n-2K_j)!}{[n - 2(K_j+k_j)]!}}
\end{equation}
with $K_j$ as in Eq.~(\ref{KJ}). Given $l$ vertices on the left side of the diagram, we have that the pertaining elementary process creates $K_l$ SH photons. Similarly, one has the same expressions for the factors and the number of generated SH photons given by the $l'$ vertices on the left side.
\\
Ultimately, we find that a double-sided diagram with $l$ vertices on the left side and $l'$ on the right side contributes to the order $R=r+r'$ of the SH photon-number distribution with a term
\begin{equation} \label{gencontr}
    \begin{aligned}
       &\text{diag}(r,r',l,l',\gamma) = \frac{i^r(-i)^{r'}}{r!r'!}\gamma^R \prod_{a=1}^{(l+1)/2}f_{\text{shc}}(n,k_{2a-1})\prod_{b=1}^{(l'+1)/2}f_{\text{shc}}(n,k_{2b-1}') \\
       &\prod_{c=1}^{(l-1)/2}f_{\text{sha}}(n,k_{2c})\prod_{d=1}^{(l'-1)/2}f_{\text{sha}}(n,k_{2d}')|n-2K_l,K_l\rangle \langle n-2K_{l'},K_{l'}|
    \end{aligned}
\end{equation}
and we get a non-null contribution only if $K_l = K_{l'}$. In the case $\mathbf{k} = \mathbf{k'}$, Eq.~(\ref{gencontr}) simplifies to
\begin{equation}
\begin{aligned}
   \text{diag}(r=r',l=l',\gamma) =& \left(\frac{\gamma^r}{r!}\right)^2\,\prod_{a=1}^{(l+1)/2}f^2_{\text{shc}}(n,k_{2a-1})\prod_{c=1}^{(l-1)/2}f^2_{\text{sha}}(n,k_{2c}) \\
   &|n-2K_l,K_l\rangle \langle n-2K_{l},K_{l}|
\end{aligned}
\end{equation}
and provides the probability of generating $K_l$ SH photons through the elementary process $A_{\mathbf{k}}$ in the distribution expansion.

Finally, we apply the procedure described above to a non-trivial process superposition to retrieve the corresponding contribution $\text{diag}(r,r',l,l',\gamma)$. 
\\
Let us find the term of the distribution expansion yielded by the superposition of the two processes $A_{\mathbf{k}}$ and $A_{\mathbf{k'}}$ with $\mathbf{k}=(2,1,4,1)$ and $\mathbf{k'} = (4)$. The corresponding double-sided diagrams are given by
\\
\\
\\
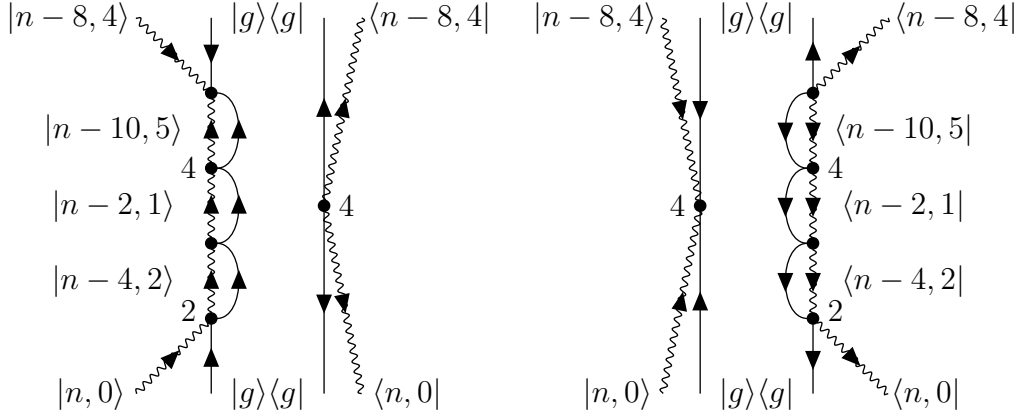
\begin{tikzpicture}
\begin{feynhand}    
\vertex (a) at (0,0);      
\vertex (b) at (1,0);
\vertex [dot] (c) at (1,1) {};   
\vertex [dot] (d) at (1,2) {};
\vertex [dot] (e) at (1,3) {};
\vertex [dot] (f) at (1,4) {};
\vertex (g) at (1,5);
\vertex (h) at (0,5);
\propag [chabos] (a) to (c);
\propag [fer] (b) to (c);
\propag [chabos] (c) to (d);
\propag [fer] (c) to [out=0, in=0] (d);
\propag [chabos] (d) to (e);
\propag [fer] (d) to [out=0, in=0] (e);
\propag [chabos] (e) to (f);
\propag [fer] (e) to [out=0, in=0] (f);
\propag [fer] (g) to (f);
\propag [chabos] (h) to (f);
\node at (-0.9,5) {$|n-8,4\rangle$}; 
\node at (-0.3,3.5) {$|n-10,5\rangle$}; 
\node at (-0.3,2.5) {$|n-2,1\rangle$}; 
\node at (-0.3,1.5) {$|n-4,2\rangle$};
\node at (-0.6,0) {$|n,0\rangle$};
\node at (1.5,0) {$|g\rangle$}; 
\node at (1.5,5) {$|g\rangle$}; 
\node at (0.7,1.1) {2};
\node at (0.7,3) {4};
\vertex (a') at (2.5,0);      
\vertex (b') at (3,0);
\vertex [dot] (c') at (2.5,2.5) {};   
\vertex (d') at (2.5,5);
\vertex (e') at (3,5);
\propag [fer] (c') to (a');
\propag [chabos] (c') to (b');
\propag [fer] (c') to (d');
\propag [chabos] (c') to (e');
\node at (2,0) {$\langle g|$};
\node at (2,5) {$\langle g|$};
\node at (3.6,0) {$\langle n,0|$};
\node at (3.9,5) {$\langle n-8,4|$};
\node at (2.8,2.5) {4};

\vertex (a') at (7,0);      
\vertex (b') at (7.5,0);
\vertex [dot] (c') at (7.5,2.5) {};   
\vertex (d') at (7.5,5);
\vertex (e') at (7,5);
\propag [chabos] (a') to (c');
\propag [fer] (b') to (c');
\propag [chabos] (e') to (c');
\propag [fer] (d') to (c');
\node at (8,0) {$|g\rangle$};
\node at (8,5) {$|g\rangle$};
\node at (6.4,0) {$|n,0\rangle$};
\node at (6.1,5) {$|n-8,4\rangle$};
\node at (7.2,2.5) {4};
\vertex (a) at (9,0);      
\vertex (b) at (10,0);
\vertex [dot] (c) at (9,1) {};   
\vertex [dot] (d) at (9,2) {};
\vertex [dot] (e) at (9,3) {};
\vertex [dot] (f) at (9,4) {};
\vertex (g) at (9,5);
\vertex (h) at (10,5);
\propag [fer] (c) to (a);
\propag [chabos] (c) to (b);
\propag [chabos] (d) to (c);
\propag [fer] (d) to [out=180, in=180] (c);
\propag [chabos] (e) to (d);
\propag [fer] (e) to [out=180, in=180] (d);
\propag [chabos] (f) to (e);
\propag [fer] (f) to [out=180, in=180] (e);
\propag [fer] (f) to (g);
\propag [chabos] (f) to (h);
\node at (8.5,5) {$\langle g|$}; 
\node at (8.5,0) {$\langle g|$};
\node at (10.9,5) {$\langle n-8,4|$};
\node at (10.2,3.5) {$\langle n-10,5|$}; 
\node at (10.2,2.5) {$\langle n-2,1|$}; 
\node at (10.2,1.5) {$\langle n-4,2|$}; 
\node at (10.5,0) {$\langle n,0|$}; 
\node at (9.3,1.1) {2};
\node at (9.3,3) {4};
\end{feynhand}
\end{tikzpicture}
\\
\\
\\
where the one on the left represents $A_{\mathbf{k}}|n,0\rangle\langle n,0|A^{\dagger}_{\mathbf{k'}}$ and the one on the right $A_{\mathbf{k'}}|n,0\rangle\langle n,0|A^{\dagger}_{\mathbf{k}}$. We know that they provide the same contribution to the SH photon-number distribution. We compute the term given by the diagram on the left.
\\
We note that for each vertex the number of annihilation operators is never larger than the number of creation operators and that the number of SH photons created is the same from both the elementary processes. Therefore, this diagram provides a non-null contribution to the distribution. In particular, four photons are created from this process superposition. 
\\
We have $l=4$ on the left side and $l'=1$ on the right side. The corresponding perturbative orders are $r = \sum_{j=1}^l k_j = 8$ and $r'=\sum_{j=1}^{l'} k_j'=4$, respectively, yielding a contribution to the order $R = r+r' = 12$ to the SH photon-number distribution.
\\
By exploiting Eqs.~(\ref{fshc}),~(\ref{fsha}) and~(\ref{gencontr}), it is just a matter of calculations to find the perturbative term given by the diagram, which reads
\begin{equation}
    \text{diag}(r=8,r'=4,l=4,l'=1,\gamma) = \frac{\gamma^{12}}{2016}\cdot\frac{n!(n-2)!}{(n-4)!(n-10)!}.
\end{equation}


\section{Conclusion}
\label{4}
We considered the Taylor expansion of the SHG unitary evolution and identified the pertaining perturbative terms as elementary processes. We found that the output SH photon-number statistics results from the superpositions of these elementary processes. We remark that these findings can be directly generalized to more complex nonlinear phenomena, such as sum- or difference-frequency generation and higher-order nonlinear effects. Moreover, the process-superposition map itself may be exploited in more general scenarios, where the elementary processes are not necessarily defined by sequences of annihilation and creation operators.
\\
We also provided a diagrammatic approach to process superpositions which straightforwardly allows one to compute their corresponding perturbative contribution to the SH photon-number distribution. This tool provides the analytical expression of high-order terms of the distribution expansion. Hence, one can further investigate the nonclassicality of SH light, the origin and the conditions for the sub-Poissonian output statistics and for the generation of squeezing.

 \bibliographystyle{elsarticle-num} 
 \bibliography{cas-refs}





\end{document}